\documentclass[reprint,
superscriptaddress,
 amsmath,amssymb,
 aps,
]{revtex4-2}

\usepackage{graphicx}
\usepackage{dcolumn}
\usepackage{bm}
\usepackage{xcolor}
\usepackage{placeins}
\usepackage{tikz}
\usepackage{cleveref}

\begin{document}


\title{Magnetic Diffuse Scattering Study of Dipolar-Octupolar Pyrochlore Ce$_2$Sn$_2$O$_7$}
\author{B.~Yuan}
\affiliation{Department of Physics and Astronomy, McMaster University, Hamilton, Ontario L8S 4M1, Canada}

\author{M. Powell}
\affiliation{Department of Chemistry, Clemson University, Clemson, South Carolina 29634-0973, USA}

\author{X.~Liu}
\affiliation{Department of Physics and Astronomy, McMaster University, Hamilton, Ontario L8S 4M1, Canada}

\author{J. Ni}
\affiliation{Department of Physics and Astronomy, McMaster University, Hamilton, Ontario L8S 4M1, Canada}

\author{E. M. Smith}
\affiliation{Department of Physics and Astronomy, McMaster University, Hamilton, Ontario L8S 4M1, Canada}

\author{R Sch\"afer}
\affiliation{Department of Physics, Harvard University, Cambridge, MA 02138, USA}

\author{R. Moessner}
\affiliation{Max Planck Institute for Physics of Complex Systems, Nöthnitzer Strasse 38, Dresden 01187, Germany}

\author{F. Ye}
\affiliation{Neutron Scattering Division, Oak Ridge National Laboratory, Oak Ridge, Tennessee 37831, USA}

\author{J.~Dudemaine}
\affiliation{Département de Physique, Université de Montréal, Montréal, Québec, Canada}
\affiliation{Regroupement Québécois sur les Matériaux de Pointe (RQMP)}
\author{A. D. Bianchi}
\affiliation{Département de Physique, Université de Montréal, Montréal, Québec, Canada}
\affiliation{Regroupement Québécois sur les Matériaux de Pointe (RQMP)}
\affiliation{Institut Courtois, Complexe des sciences, Université de Montréal,
1375 Ave. Thérèse-Lavoie-Roux, Montréal, Québec H2V 0B3, Canada}

\author{J. W. Kolis}
\affiliation{Department of Chemistry, Clemson University, Clemson, South Carolina 29634-0973, USA}

\author{B.~D.~Gaulin}
\affiliation{Department of Physics and Astronomy, McMaster University, Hamilton, Ontario L8S 4M1, Canada}
\affiliation{Brockhouse Institute for Materials Research, McMaster University, Hamilton, Ontario L8S 4M1, Canada}
\affiliation{Canadian Institute for Advanced Research, 661 University Avenue, Toronto, Ontario M5G 1M1, Canada.}
\date{\today}
\begin{abstract}
We report diffuse scattering measurements on the dipolar-octupolar pyrochlore, Ce$_2$Sn$_2$O$_7$, to elucidate its magnetic correlations within the correlated paramagnetic regime above its ordering temperature, $\mathrm{T_N}\sim 40~$mK. The magnetic transition at $\mathrm{T_N}$, observed in recent specific heat measurements, has been interpreted within the nearest-neighbor XYZ model as a transition into an all-in-all-out (AIAO) magnetic order. However, the observed diffuse scattering above $\mathrm{T_N}$ argues against an AIAO magnetic order by exhibiting $no$ short-range order at the AIAO wave-vectors, inconsistent with the nearest neighbor model. Instead, the observed diffuse scattering strongly resembles that observed in classical dipolar spin ice, Dy$_2$Ti$_2$O$_7$, and suggests the presence of short-range correlated structures such as chains, or clusters in the low energy manifold of Ce$_2$Sn$_2$O$_7$, favored by further neighbor interactions. We discuss how chain-like states qualitatively account for the observed H-T phase diagram under a [1,1,0]-field.
\end{abstract}
\maketitle
\section{Introduction}
Searching for an experimental realization of quantum spin ice (QSI) has been a central theme in the study of rare-earth (RE) pyrochlores \citep{Balents_PRB_2004, Gingras_review_2014}. Over nearly two decades of intensive experimental efforts, a specific class known as the dipolar-octupolar (DO) pyrochlores \citep{PRL_Gangchen2014, JeffRau_review, DOreview}, such as those based on Ce$^{3+}$ and Nd$^{3+}$, has emerged as the leading QSI candidates. Central to their magnetism are the unique symmetry properties of their Kramers' doublet ground state wave-functions that give rise to the unusual transformation properties of different effective pseudo-spin components, $S_{\alpha=x,y,z}$ ($\hat{z}$ is parallel to the local 3-fold axis), under the local $D_{3d}$ point group and time-reversal operations. Specifically, only the $S_x$ and $S_z$ components transform like a magnetic dipole (although only $S_z$ carries a physical dipole moment that couples linearly to magnetic field and neutron spin), whereas $S_y$ transforms like a magnetic octupole \citep{PRL_Gangchen2014, JeffRau_review, DOreview}. In contrast to the complex generic nearest neighbor (NN) interactions in RE-pyrochlores based on usual Kramers' ions with only dipolar components (e.g. Er$^{3+}$ and Yb$^{3+}$) \citep{JeffRau_review}, the NN interaction between DO pseudo-spins is constrained to take the following simple form:
\begin{align}\label{XYZ}
J_{\tilde{x}}S_{\tilde{x},i}S_{\tilde{x},j}+J_{\tilde{y}}S_{\tilde{y},i}S_{\tilde{y},j}+J_{\tilde{z}}S_{\tilde{z},i}S_{\tilde{z},j} 
\end{align} 
in a rotated spin coordinate given by $S_{\tilde{x}/\tilde{z}}=\cos(\theta)S_{x/z}\pm\sin(\theta)S_{z/x}$ and $S_{\tilde{y}}=S_y$, where $\theta$ is a material-dependent parameter \citep{PRL_Gangchen2014, JeffRau_review, DOreview}.

The simplicity of the phase diagram associated with Eq.~\ref{XYZ}, featuring an extended QSI phase alongside a $\mathbf{Q}=0$ all-in-all-out (AIAO) order \citep{OwenBenton_PRB_2020, YBKim_PRL_2024}, has prompted extensive studies of Nd-based and Ce-based pyrochlores. In particular, the NN XYZ model has been found to provide an excellent description of all the Nd-pyrochlores studied so far \citep{BellaLake_NatPhys_2016_Nd2Zr2O7, OwenBenton_PRB_2016_Nd2Zr2O7, BellaLake_PRB_2019_Nd2Zr2O7, PRL_BellaLake_2020_Nd2Zr2O7, SPetit_PRL_2021_Nd2Zr2O7, AdamAczel_2025_Arxiv_Nd2Sn2O7}. Specifically, modeling the high temperature thermodynamic properties, and the spin-wave excitations in these systems finds competing $J_{\tilde{x}}>0$ and $J_{\tilde{z}}<0$ \citep{BellaLake_NatPhys_2016_Nd2Zr2O7, OwenBenton_PRB_2016_Nd2Zr2O7, AdamAczel_2025_Arxiv_Nd2Sn2O7}, between which the smaller but unfrustrated $J_{\tilde{z}}$ eventually gives rise to an AIAO order.

More recently, the Ce-pyrochlores, Ce$_2$X$_2$O$_7$ (X=Sn, Hf, Zr) \citep{Sibille_PRL_2015_Ce2Sn2O7, Natphys_Pengchengdai_2019_Ce2Zr2O7,PRL_gaudet_2019_Ce2Zr2O7, Sibille_NatPhys_2020_Ce2Sn2O7,PhysRevB.106.094425,Bhardwaj2022_npjquantummaterials, EvanSmith_PRX_2022_Ce2Zr2O7, EvanSmith_PRB_Ce2Zr2O7_2023, JamesBeare_PRB_Ce2Zr2O7_2023, Yahne_PRX_2024_Ce2Sn2O7, Sibille_PRB_2025_Ce2Hf2O7, EvanSmith_PRL_2025_Ce2Hf2O7, EvanSmith_PRX_2025_Ce2Zr2O7, Kermarrec_Ce2Hf2O7_2025_arxiv, Sibille_NatPhys_2025_Ce2Sn2O7, NatPhys_Pengchengdai_2025_Ce2Zr2O7, gao2026spectroscopicdemarcationemergentphotons} have attracted much attention for displaying QSI-like behaviors such as a broad `snow-flake' like diffuse scattering pattern, as well as scattering continua in their excitation spectra. Although most of the existing data are still interpreted within the same theoretical framework based on the NN XYZ model \citep{Sibille_NatPhys_2020_Ce2Sn2O7, PhysRevB.106.094425, EvanSmith_PRX_2022_Ce2Zr2O7,Yahne_PRX_2024_Ce2Sn2O7, Sibille_PRB_2025_Ce2Hf2O7, EvanSmith_PRL_2025_Ce2Hf2O7,Kermarrec_Ce2Hf2O7_2025_arxiv, Sibille_NatPhys_2025_Ce2Sn2O7, NatPhys_Pengchengdai_2025_Ce2Zr2O7, gao2026spectroscopicdemarcationemergentphotons}, deviation from this simple model has been inferred from the specific heat measurements on both Ce$_2$Zr$_2$O$_7$ \citep{EvanSmith_PRX_2022_Ce2Zr2O7, Bhardwaj2022_npjquantummaterials} and Ce$_2$Hf$_2$O$_7$ \citep{EvanSmith_PRL_2025_Ce2Hf2O7} - where the calculated and measured specific heat diverge below a moderate temperature of $\sim 0.5~\mathrm{K}$ \citep{EvanSmith_PRX_2022_Ce2Zr2O7} and $\sim 0.25~\mathrm{K}$ \citep{EvanSmith_PRL_2025_Ce2Hf2O7}, respectively, still above the convergence limit of the numerical linked clusters (NLC) method used. Since the XYZ interaction constitutes the \textit{only} symmetry allowed terms between nearest neighbors, these discrepancies suggest the presence of interactions beyond the NN model \citep{EvanSmith_PRX_2022_Ce2Zr2O7, EvanSmith_PRL_2025_Ce2Hf2O7}. However, attempts to elucidate their roles are hampered by the presence of structural disorder in samples grown by conventional solid-state and floating zone methods, manifested as significant sample-to-sample variability of the specific heat of both Ce$_2$Hf$_2$O$_7$ \citep{EvanSmith_PRL_2025_Ce2Hf2O7, Sibille_PRB_2025_Ce2Hf2O7} and Ce$_2$Zr$_2$O$_7$ \citep{EvanSmith_PRX_2022_Ce2Zr2O7, Natphys_Pengchengdai_2019_Ce2Zr2O7, POPA200870}. Consequently, the following critical question still remains: to what extent is the \textit{intrinsic} low energy physics in the Ce-pyrochlores described by the simple NN model, and what roles do further neighbor terms - inevitable in real materials - play in their microscopic spin correlations?

Recent advances in hydrothermal synthesis of the last member of the Ce$_2$X$_2$O$_7$ family, Ce$_2$Sn$_2$O$_7$, provide a unique opportunity to address the above question \citep{hydrothermalgrowth}. Unlike conventional solid-state and floating-zone growth methods, carried out at temperatures above $1000^{\circ}\mathrm{C}$ and $2000^{\circ}\mathrm{C}$ for powder and single-crystal synthesis, respectively, this technique requires a substantially lower temperature of $700^{\circ}\mathrm{C}$. Extensive earlier studies on sample dependence in rare-earth pyrochlores, particularly in Yb$_2$Ti$_2$O$_7$ \citep{PhysRevB.84.172408, Yb2Ti2O7sample, Shafieizadeh2018}, have demonstrated that a low synthesis temperature is essential for minimizing structural disorder and revealing their intrinsic low-temperature physics. Consistent with this, our hydrothermally grown Ce$_2$Sn$_2$O$_7$ crystals show no measurable nuclear diffuse scattering indicating a very low level, if not an absence, of disorder, in contrast to our previous floating-zone-grown Ce$_2$Zr$_2$O$_7$ crystals (see Fig.~\ref{Ce2Zr2O7_defect} and Fig.~\ref{Ce2Zr2O7_Pr2Zr2O7}).

By \textit{simultaneously} modelling neutron powder diffraction (NPD), specific heat and magnetic susceptibility measurements, our previous work on hydrothermally grown Ce$_2$Sn$_2$O$_7$ powder uniquely constrains the exchange parameters within the framework of the NN XYZ model \citep{Yahne_PRX_2024_Ce2Sn2O7}. Specifically, quantitative agreement with experiments is achieved over a considerably wider temperature range (down to at least $\sim$0.1~K) than analogous studies on Ce$_2$Zr$_2$O$_7$ and Ce$_2$Hf$_2$O$_7$, finding 
\begin{align}\label{parameters}
\{J_{\tilde{x}},J_{\tilde{y}},J_{\tilde{z}}\}&=\{0.0450,-0.001,-0.012\}\mathrm{meV} \notag \\
\theta& =0.19\pi     
\end{align}. These parameters predict a transition to AIAO order at $\mathrm{T_N}\sim 50~\mathrm{mK}$. Remarkably, a sharp transition at $\mathrm{T_N\sim 40~mK}$ was in fact observed in our recent specific heat measurement on hydrothermally grown Ce$_2$Sn$_2$O$_7$ \citep{smith2026zerofieldlongrangeorder}. However, despite the good agreement between the calculated and observed value of $\mathrm{T_N}$ at zero field, the H-T phase diagram under a [1,1,0]-field is incompatible with an AIAO ground state in two ways. First, the zero-field magnetic order is \textit{stabilized} by an external magnetic field, as evidenced by an \textit{enhancement of} $\mathrm{T_N}$ from $\sim 40~$mK at 0~T to $\sim 90~$mK at $\sim 2~$T. This is inconsistent with the expected behavior of AIAO order, which is antiferromagnetic and therefore should be suppressed by field \citep{BellaLake_PRB_2019_Nd2Zr2O7, Nd2Zr2O7fielddep,Nd2Zr2O7fielddep2}. Furthermore, the pyrochlore lattice under a [1,1,0]-field is decomposed into \textit{ferromagnetic} $\alpha-$ and $\beta-$chains along [1,1,0] and [1,-1,0], respectively \citep{Nd2Zr2O7fielddep2, EvanSmith_PRB_Ce2Zr2O7_2023,smith2026zerofieldlongrangeorder}. The zero-field magnetic order seems to evolve smoothly to such a field-induced chain-like phase with no additional phase boundary between these two phases, an observation that is hard to reconcile with an AIAO order at 0~T \citep{BellaLake_PRB_2019_Nd2Zr2O7, Nd2Zr2O7fielddep, Nd2Zr2O7fielddep2}. Addressing these questions and assessing the validity of the NN model call for a careful study of magnetic correlations in Ce$_2$Sn$_2$O$_7$.

In this paper, we report diffuse scattering measurements on a co-aligned array of hydrothermally grown Ce$_2$Sn$_2$O$_7$ single crystals in the correlated paramagnetic regime at 50~mK and 800~mK above $\mathrm{T_N}$. Importantly, we observed no short-range order associated with the AIAO order, which was predicted by the nearest neighbor XYZ model to be present over a large temperature range above $\mathrm{T_N}$. Our results therefore argue against AIAO as the ground state magnetic order in Ce$_2$Sn$_2$O$_7$, thereby challenging the validity of the NN model given by Eq.~\ref{XYZ} and Eq.~\ref{parameters}. Remarkably, the resulting spin correlations in Ce$_2$Sn$_2$O$_7$ are almost identical to those observed in \textit{classical} dipolar spin ice \citep{Fennell_PRB_2004, Samarakoon2020}, suggesting the relevance of short-range correlated structures, such as chains or clusters in Ce$_2$Sn$_2$O$_7$. We discuss how the presence of chain-like states provides a natural explanation for the H-T phase diagram for a [1,1,0]-field.

\section{Experimental Details}
\begin{figure}[tb]
\includegraphics[width=0.5\textwidth]{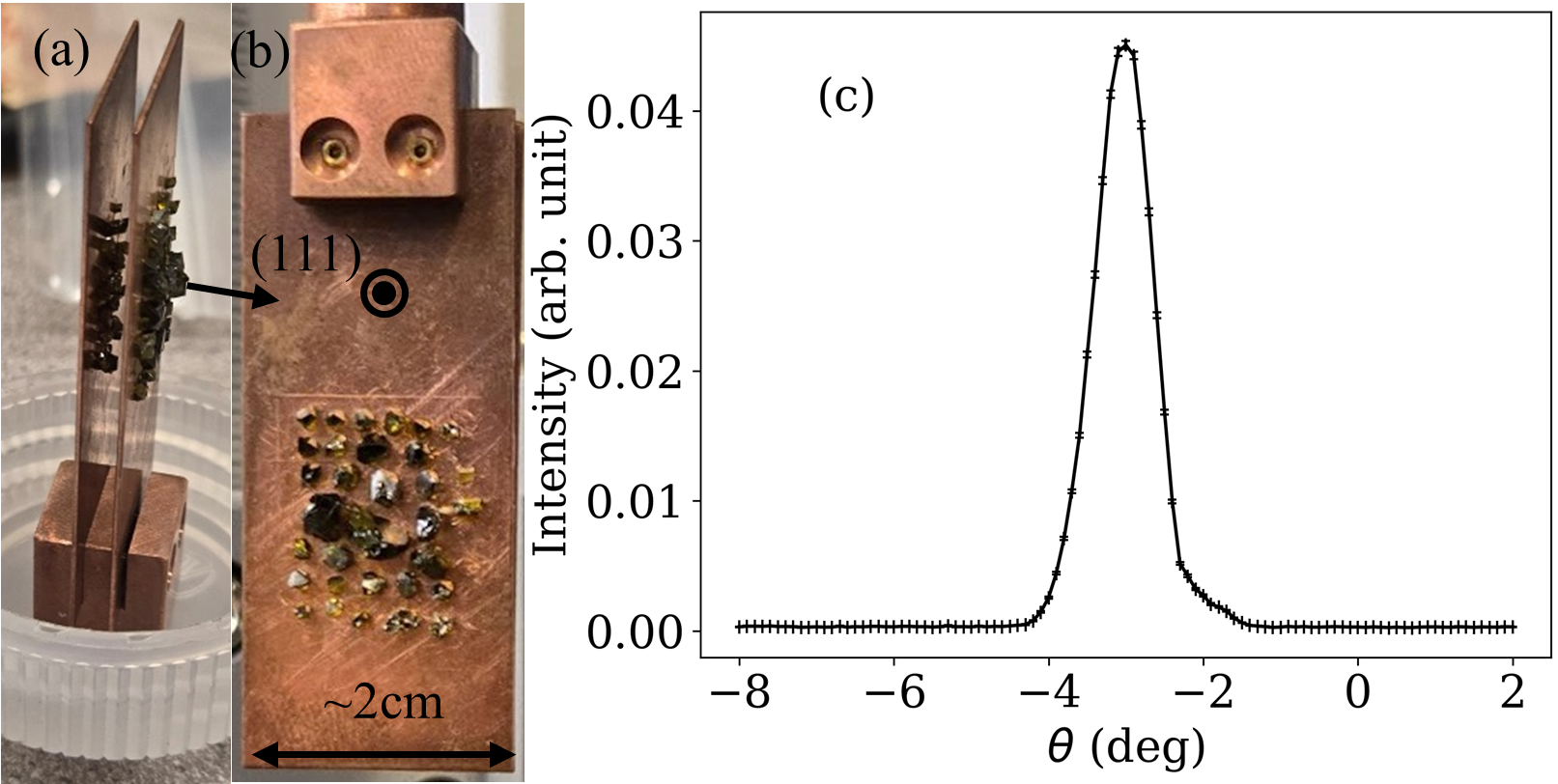}
\caption{\label{sample} (a) Side and (b) front view of the array of co-aligned Ce$_2$Sn$_2$O$_7$ single crystals used in the diffuse scattering experiment. (111) is perpendicular to the copper plates. (c) Rocking scan around the (440) nuclear Bragg peak of the co-aligned single crystal array.}
\end{figure}

Diffuse neutron scattering measurements were carried out on a co-aligned array of hydrothermally grown Ce$_2$Sn$_2$O$_7$ single crystals with a total mass of $\sim 0.8~g$. Details of the hydrothermal growth have been reported elsewhere~\citep{hydrothermalgrowth}. As shown in Fig.~\ref{sample}~(a, b), 55 small single crystals (the largest one weighs $\sim$ 125~mg and is $\sim$5~mm across), co-aligned using an X-ray Laue diffractometer, are glued to two 0.6~mm thick copper plates using GE-varnish. The effective mosaicity of the co-aligned single-crystal array is confirmed to be $\sim 1$ degree on the McMaster All-purpose Diffractometer (MAD) at the McMaster Nuclear Reactor [Fig~\ref{sample}(c)]. 

Diffuse neutron scattering measurements were carried out at mixing chamber temperatures of 50~mK, 800~mK and 12~K using a dilution refrigerator on the CORELLI spectrometer at the Spallation Neutron Source (SNS), Oak Ridge National Laboratory (ORNL). The single-crystal array was mounted with [H,H,L] coincident with the horizontal plane, and was rotated over 270 degrees with a 1 degree step at each temperature. Throughout the experiment, a correlation chopper rotating at a fixed frequency of 293.412~Hz was employed before the sample to partially reject the high energy inelastic background. Only data after cross-correlation, which integrates over an energy window $\Delta \hbar\omega \approx 0.05 \mathrm{E_i} \approx 0.4-2.5~\mathrm{meV}$ \citep{CORELLI_instrument}, are presented in the main text and the Supplemental Materials. Given the low magnetic energy scale ($< 0.1~$meV) in the Ce-pyrochlores, the resulting diffuse scattering data can be regarded as integrating over all magnetic fluctuations in Ce$_2$Sn$_2$O$_7$.

\begin{figure*}[tb]
\includegraphics[width=1\textwidth]{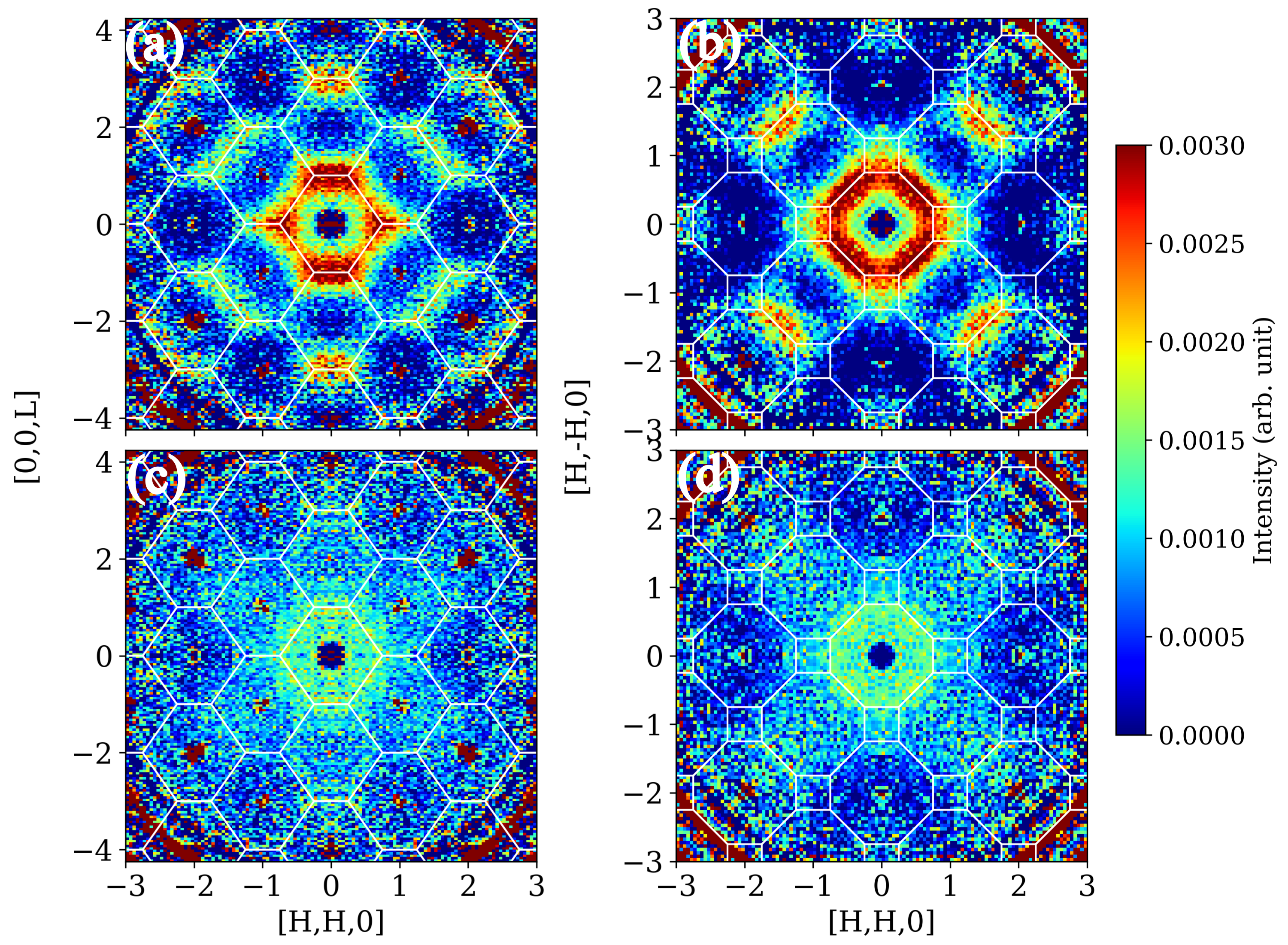}
\caption{\label{Ce2Sn2O7large} Ce$_2$Sn$_2$O$_7$ diffuse neutron scattering in the (a,c) HHL and (b,d) HK0 plane at (a,b) 50~mK and (c,d) 800~mK. A 12~K background has been subtracted from all the data. The data in (a,c)/(b,d) is obtained by integrating the diffuse intensity in the [H,-H,0]/[0,0,L] direction with $-0.2<\mathrm{H}<0.2$/$-0.2<\mathrm{L}<0.2$, and has been symmetrized in accordance with the $Fd\Bar{3}m$ space group. The white lines denote the Brillouin zone boundaries of the primitive FCC lattice.}
\end{figure*}

\section{Results}
Energy-integrated neutron intensity maps at 50~mK in the HHL [Fig.~\ref{Ce2Sn2O7large}(a)] and HK0 plane [Fig.~\ref{Ce2Sn2O7large}(b)] (a 12~K background has been subtracted from all data in Fig.~\ref{Ce2Sn2O7large}) reveal a highly structured 3-dimensional diffuse scattering pattern in Ce$_2$Sn$_2$O$_7$. The pattern essentially disappears at 800~mK [Fig.~\ref{Ce2Sn2O7large}(c, d)], confirming its magnetic origin. Our data clearly show that the diffuse scattering in Ce$_2$Sn$_2$O$_7$ is concentrated along the Brillouin zone (BZ) boundaries of the underlying FCC lattice (indicated by white solid lines in Fig.~\ref{Ce2Sn2O7large}). The intensity is maximized in the first BZ and weakens at larger $|\mathbf{Q}|$, consistent with the expected behavior of the dipolar form factor \citep{EvanSmith_PRX_2025_Ce2Zr2O7}. On the other hand, no magnetic Bragg peak or diffuse scattering intensity is observed at the AIAO ordering wave-vectors such as (220) and (113) (See also Supplemental Materials). The data presented in Fig.~\ref{Ce2Sn2O7large} are of markedly better quality than previous measurements on Ce$_2$Zr$_2$O$_7$ \citep{PRL_gaudet_2019_Ce2Zr2O7, Natphys_Pengchengdai_2019_Ce2Zr2O7} and Ce$_2$Hf$_2$O$_7$ \citep{EvanSmith_PRL_2025_Ce2Hf2O7}, which observed a ring-like diffuse signal along the boundary of the first BZ qualitatively similar to the present data, but only a broad $\mathbf{Q}$-dependence beyond the first zone. In contrast, the diffuse scattering in Ce$_2$Sn$_2$O$_7$ remains well-defined and unambiguously resides along the FCC BZ boundaries even in the second and third BZs [e.g. around (222) in Fig.~\ref{Ce2Sn2O7large}(a)]. In addition to a dramatic improvement in crystal quality, a key to the exceptional data quality in Fig.~\ref{Ce2Sn2O7large} is the large reciprocal space coverage at CORELLI (enabled by the use of thermal neutrons and a large detector coverage in the horizontal and vertical directions), which enables us to probe \textit{all} symmetry equivalent directions in the full \textit{3-dimensional} reciprocal space. Folding the data by $all$ symmetry operations of the cubic space group significantly improves the signal-to-noise ratio, although we emphasize that the observed diffuse scattering pattern is already clearly visible in the raw background-subtracted data. (See Supplemental Materials).

\begin{figure}[tb]
\includegraphics[width=0.5\textwidth]{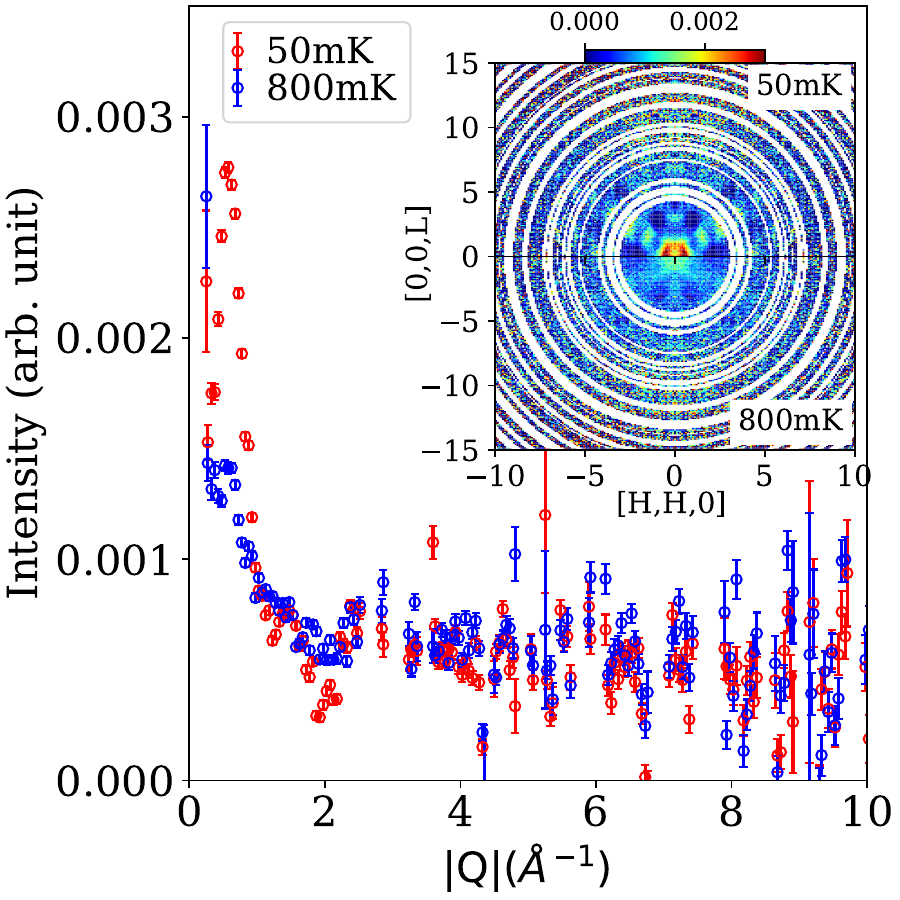}
\caption{\label{largeQ} Powder-averaged diffuse scattering intensity as a function of $|\mathbf{Q}|$ at 50~mK and 800~mK obtained by performing an orientational average of the symmetrized single crystal data in the [H+0.5,H-0.5,L] plane (inset). To avoid the Ce$_2$Sn$_2$O$_7$ nuclear Bragg peaks, we used an integration range of $0.3<\mathrm{H}<0.7$ in the [H,-H,0] direction. A 12~K background has been subtracted from both the single crystal and powder-averaged data. Powder rings due to aluminum sample can of the dilution insert and the copper sample mount have also been masked out in the data.}
\end{figure}

The large reciprocal space coverage also enabled us to study the magnetic correlations in Ce$_2$Sn$_2$O$_7$ at large $|\mathbf{Q}|$. Following our previous analysis of the Ce$_2$Zr$_2$O$_7$ CORELLI data \citep{EvanSmith_PRX_2025_Ce2Zr2O7}, we examined the diffuse scattering in the [H+0.5,H-0.5,L] plane (inset of Fig.~\ref{largeQ}) which contains no nuclear Bragg peaks (powder rings due to the sample environment and the sample holder have also been removed). Other than the zone boundary (ZB) scattering at low $|\mathbf{Q}|$'s, our data shows no obvious magnetic scattering beyond $|\mathbf{Q}|\sim 3\AA^{-1}$ (or $\mathrm{L}\sim5$). To directly compare with the previous NPD measurements, we performed an orientational average of the single crystal data in Fig.~\ref{largeQ} to obtain the diffuse scattering intensity as a function of $|\mathbf{Q}|$. The resulting spectrum (main panel of Fig.~\ref{largeQ}) consists of a strong peak at $|\mathbf{Q}|\sim 0.6~\AA^{-1}$ and a weak peak at $|\mathbf{Q}|\sim 1.4~\AA^{-1}$ arising from the ZB scattering around the first and second BZs, respectively, followed by a temperature-independent flat background at $|\mathbf{Q}|\gtrsim 3~\AA^{-1}$. As shown in Fig.~\ref{powdercompare}, the orientationally averaged data is in excellent agreement with the previous NPD results on hydrothermally grown Ce$_2$Sn$_2$O$_7$ powder \citep{Yahne_PRX_2024_Ce2Sn2O7}. The data in Fig.~\ref{largeQ} and Ref.~\citep{Yahne_PRX_2024_Ce2Sn2O7} are to be contrasted with NPD on powder samples grown by conventional solid-state methods that show no low $\mathbf{Q}$ dipolar scattering, but only a broad peak centered at $|\mathbf{Q}|\sim 8~\AA^{-1}$ previously attributed to octupolar scattering \citep{Sibille_NatPhys_2020_Ce2Sn2O7}. The apparent absence of any octupolar scattering in Ce$_2$Sn$_2$O$_7$, and previously in Ce$_2$Zr$_2$O$_7$, has been argued to be a simple consequence of the much larger Ce$^{3+}$ dipolar form factor compared to its octupolar form factor which alone makes the scattering by the $S_z$-component approximately 50 times stronger than that by the $S_x/S_y$ components \citep{EvanSmith_PRX_2025_Ce2Zr2O7}. Consequently, in an energy-integrated unpolarized neutron diffraction experiment that detects fluctuations in all pseudo-spin components, the diffuse scattering intensity is overwhelmingly dominated by $S_z$ following the dipolar form factor, largely independent of the underlying spin model \citep{EvanSmith_PRX_2025_Ce2Zr2O7}. 

\begin{figure*}[tb]
\includegraphics[width=0.8\textwidth]{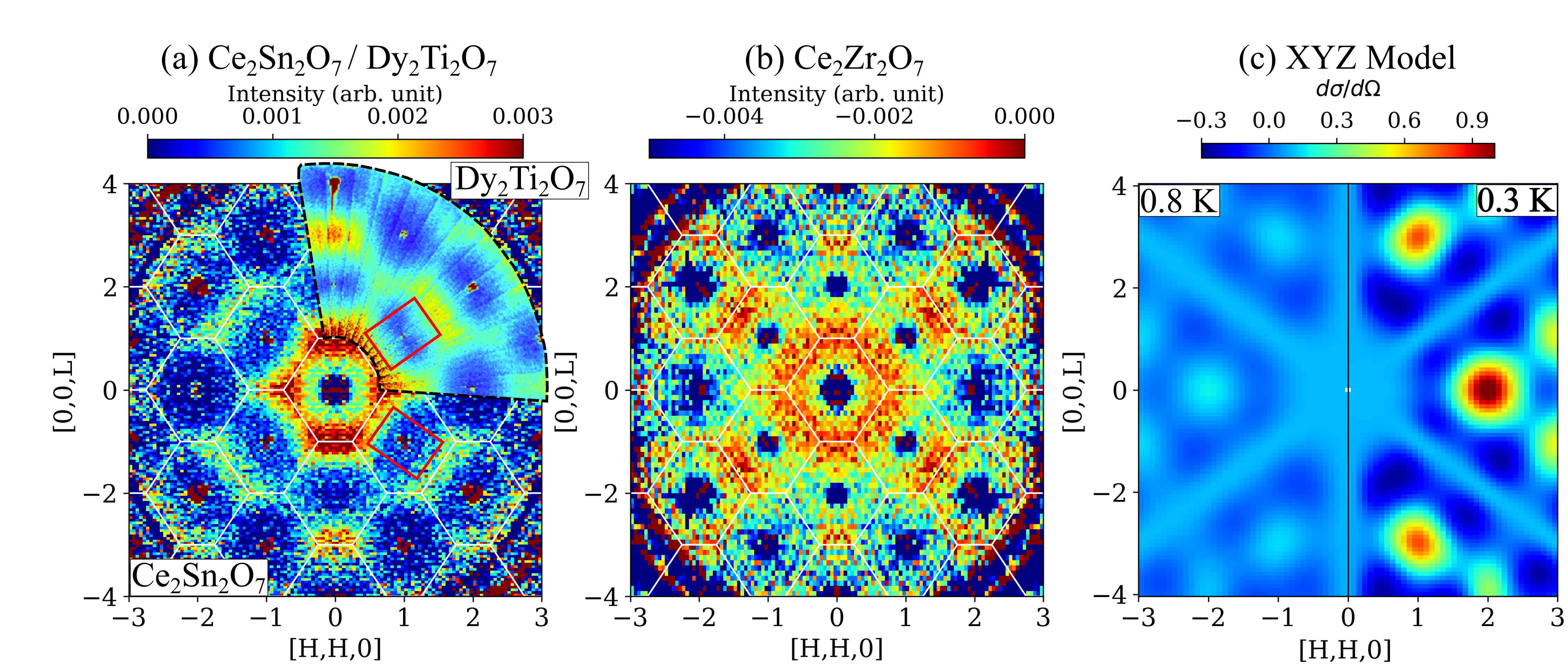}
\caption{\label{comparison} (a, b) Symmetrized diffuse neutron scattering in the [H,H,L] plane for (a) Ce$_2$Sn$_2$O$_7$ [same as Fig.~\ref{Ce2Sn2O7large}~(a)] and (b) Ce$_2$Zr$_2$O$_7$ at 50~mK. A high-temperature background has been subtracted in both panels. An unsymmetrized version of (b) was previously published in Ref.~\citep{EvanSmith_PRX_2025_Ce2Zr2O7}. Published Dy$_2$Ti$_2$O$_7$ data at 300~mK, adapted from Ref.~\citep{Fennell_PRB_2004}, are overlaid on the Ce$_2$Sn$_2$O$_7$ data in (a). (c) Diffuse scattering  at 800~mK (left) 300~mK (right), after subtracting a 12~K background, as predicted by the sixth-order NLC calculations with XYZ Hamiltonian parameters in Eq.~\ref{parameters}.}
\end{figure*}

\section{Discussion} 
\subsection{Comparison with the NN Model}
Our previous specific heat measurements confirmed the presence of magnetic order in Ce$_2$Sn$_2$O$_7$ below a $\mathrm{T_N}\sim$ 40~mK \citep{smith2026zerofieldlongrangeorder}. Although the experimentally accessible temperatures are above $\mathrm{T_N}$, thereby precluding the direct observation of magnetic Bragg peaks associated with the low temperature ordered phase, a robust prediction of the NN model using exchange parameters given by Eq.~\ref{parameters} is the presence of an extended correlated paramagnetic regime in which ice-rule-like correlations in the $S_{\tilde{x}}$ channel coexist with short-range AIAO correlations in the $S_{\tilde{z}}$ channel, favored by $J_{\tilde{x}}>0$ and $J_{\tilde{z}}<0$, respectively. Both contribute to the $S_z$-fluctuations with a finite $\theta$.  One therefore expects the diffuse scattering to consist of both a spin-ice-like pattern and broad peaks at the AIAO wave-vectors such as (220) and (113). This expectation is supported by the $6^{th}$-order NLC calculations \citep{Schafer2020, schaefer_magnetic_2022, EvanSmith_PRX_2022_Ce2Zr2O7} presented in Fig.~\ref{comparison}~(c) using the parameters in Eq.~\ref{parameters}, predicting broad peaks related to short-range AIAO order already at 800~mK. The AIAO peaks intensify upon approaching $\mathrm{T_N}$, and become the \textit{dominant} features already at 300~mK, well above our experimental temperature. Such behavior is indeed observed in Nd$_2$Zr$_2$O$_7$ \citep{SPetit_PRL_2021_Nd2Zr2O7, PRL_BellaLake_2020_Nd2Zr2O7}, where the validity of the NN XYZ model has been well established \citep{BellaLake_NatPhys_2016_Nd2Zr2O7, OwenBenton_PRB_2016_Nd2Zr2O7, BellaLake_PRB_2019_Nd2Zr2O7}. 

In contrast, no such AIAO-like correlations are discernible in our data [Fig.~\ref{comparison}~(a)], largely ruling out AIAO order as the ground state magnetic order in Ce$_2$Sn$_2$O$_7$. The exchange parameters within the NN model were previously shown to be strongly constrained by diffuse scattering, specific heat, and susceptibility measurements on polycrystalline Ce$_2$Sn$_2$O$_7$. In other words, there is little room for parameter adjustment within the NN framework. We therefore attribute the large discrepancy between the observed and predicted spin correlations above $\mathrm{T_N}$ to further-neighbor interactions beyond the NN model.

\subsection{Comparison with Dy$_2$Ti$_2$O$_7$}
\begin{figure}[tb]
\includegraphics[width=0.5\textwidth]{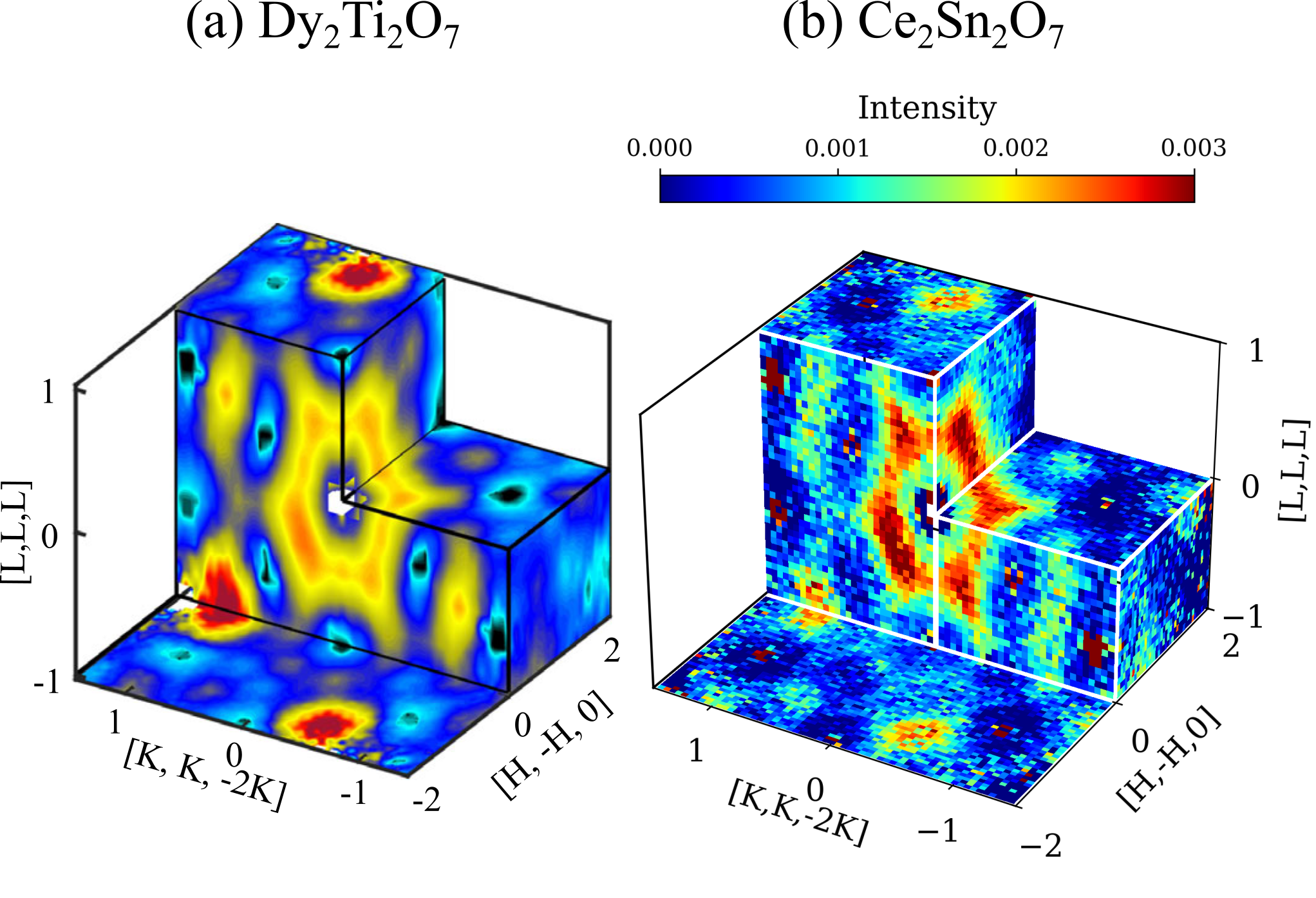}
\caption{\label{3D} Three-dimensional diffuse scattering in (a) Dy$_2$Ti$_2$O$_7$ at 680~mK, adapted from Ref.~\citep{Samarakoon2020} and (b) Ce$_2$Sn$_2$O$_7$ at 50~mK. The Ce$_2$Sn$_2$O$_7$ data have been symmetrized and background-subtracted, as in Fig.~\ref{Ce2Sn2O7large}(a) and Fig.~\ref{comparison}(a).}
\end{figure}

Insights to the nature of spin correlations in Ce$_2$Sn$_2$O$_7$ are gained by noting that its diffuse scattering is almost identical to that observed in the classical dipolar spin ice (DSI), Dy$_2$Ti$_2$O$_7$ \citep{Fennell_PRB_2004, Samarakoon2020}, also shown in Fig.~\ref{comparison}~(a). Fig.~\ref{3D} further highlights this similarity by revealing almost identical $\mathbf{Q}$ dependence of their full \textit{three-dimensional} diffuse scattering. Just like Ce$^{3+}$, Dy$^{3+}$ is also octupolar-dipolar \citep{JeffRau_review}, these observations therefore suggest the presence of very similar $S_z$ correlations in the two systems. 

The low energy physics in Dy$_2$Ti$_2$O$_7$ is now well-understood to be governed by the long-range magnetic dipolar interactions that partially lift the degeneracy of the 2-in-2-out (2I2O) manifold \citep{Melko_2004, PhysRevLett.95.217201} to favor the so-called `chain-like' states featuring ferromagnetic $S_z$-chains along the [1,1,0] and [1,-1,0] directions \citep{PhysRevB.92.094418}. Competition between various chain-based order, controlled by the residual magnetic dipolar and further neighbor exchange interactions, has been shown to account for the distribution of diffuse scattering intensity in Dy$_2$Ti$_2$O$_7$ \citep{PhysRevB.92.094418, Gingras_PRB_2016, 8xtq-tfx3}. Observation of very similar diffuse scattering in Ce$_2$Sn$_2$O$_7$ suggests these chain-like states might also be the relevant low-energy manifold in this system, and that the order below $\mathrm{T_N}\sim 40~$mK stems from ordering of the chains.

Since the high-field phase under a [1,1,0]-field also consists of chains along [1,1,0] ($\alpha-$ chains) and [1,-1,0] ($\beta-$ chains) \citep{EvanSmith_PRB_Ce2Zr2O7_2023, BellaLake_PRB_2019_Nd2Zr2O7, Nd2Zr2O7fielddep2}, one expects such a chain-based order to evolve smoothly as a function of field, thus accounting for the apparent absence of phase boundary separating the zero-field order and the high-field state. Moreover, unlike the zero-field phase, where both $\alpha-$ and $\beta-$ chains are active magnetic degrees of freedom, the [1,1,0]-field polarizes the $\alpha-$chains, partially removing the degeneracy of the low-energy manifold and relieving the frustration of the system. This naturally explains the observed enhancement of $\mathrm{T_N}$.

Alternatively, a \textit{phenomenological} description based on independent hexagonal clusters has been proposed to describe the diffuse scattering intensity in Dy$_2$Ti$_2$O$_7$ \citep{Bramwell_PRL_2008}. Such cluster-like states occur naturally in the presence of quantum fluctuations \citep{Benton_PRB_2012, PhysRevLett.131.096702}. The apparent absence of the `bow-tie' shaped scattering near the pinch points in Ce$_2$Sn$_2$O$_7$ [highlighted by the red rectangles in Fig.~\ref{comparison}~(a)] may also favor such a picture \citep{Bramwell_PRL_2008}. However, given the small Ce$^{3+}$ dipole moment ($\sim\frac{1}{10}$ of the Dy$^{3+}$ moment), unambiguously demonstrating the absence (or presence) of pinch-point scattering will likely require polarized neutron scattering measurements similar to those previously used to separate the weak pinch-point scattering from the much stronger ZB scattering in DSI, Ho$_2$Ti$_2$O$_7$ \citep{Fennell_Science_2009}.  Moreover, it is difficult to reconcile a cluster-based low-energy manifold with the observed H-T phase diagram under [1,1,0]-field.

\subsection{Comparison with Ce$_2$Zr$_2$O$_7$}
We have also revisited the diffuse scattering in Ce$_2$Zr$_2$O$_7$, part of which has been reported in Ref.~\citep{EvanSmith_PRX_2025_Ce2Zr2O7}. Compared to Ce$_2$Sn$_2$O$_7$, the raw background-subtracted Ce$_2$Zr$_2$O$_7$ data are of lower signal-to-noise ratio (See Supplemental Materials), likely due to lower sample quality, and the diffuse scattering pattern is only clearly resolved after symmetrization according to the cubic space group. The resulting data in Fig.~\ref{comparison}~(b) also show a concentration of diffuse scattering intensity along the ZBs, albeit with different $\mathbf{Q}$ and temperature dependence compared to Ce$_2$Sn$_2$O$_7$ (See Supplemental Materials). The ZB scattering around the first BZ in Fig.~\ref{comparison}~(b) is consistent with Ref.~\citep{PRL_gaudet_2019_Ce2Zr2O7, Natphys_Pengchengdai_2019_Ce2Zr2O7} while the symmetrized data also show well-defined ZB scattering in the second and third BZs that have not been observed previously.  As shown by Ref.~\citep{Bhardwaj2022_npjquantummaterials} using self-consistent Gaussian approximations, the presence of well-defined ZB scattering around higher-order BZs in Ce$_2$Zr$_2$O$_7$ could indeed arise from a larger next-nearest-neighbor interaction.

As discussed in Section~\ref{disorder}, besides magnetic diffuse scattering, we note the presence of a temperature-\textit{independent} but $\mathbf{Q}$-dependent nuclear diffuse scattering in Ce$_2$Zr$_2$O$_7$ that is absent in Ce$_2$Sn$_2$O$_7$. This confirms a higher level of \textit{correlated} structural defects  in Ce$_2$Zr$_2$O$_7$ (random structural defects should only give rise to $\mathbf{Q}$-independent nuclear diffuse scattering) that could account for its broader diffuse scattering pattern compared to Ce$_2$Sn$_2$O$_7$, or a shorter magnetic correlation length. However, whether these defects modify the nature of the spin correlations beyond just reducing its correlation length requires the synthesis of low-disorder Ce$_2$Zr$_2$O$_7$ samples.
Moreover, the diffuse scattering in Ce$_2$Zr$_2$O$_7$ was shown previously to have an inelastic contribution \citep{PRL_gaudet_2019_Ce2Zr2O7, Natphys_Pengchengdai_2019_Ce2Zr2O7, NatPhys_Pengchengdai_2025_Ce2Zr2O7}, implying a dynamical component of its spin correlations. Similar inelastic neutron scattering measurements on Ce$_2$Sn$_2$O$_7$ are desirable to further understand the differences in diffuse scattering between these two systems.
 
\section{Conclusions} 
By carrying out diffuse scattering measurement above $\mathrm{T_N}$, we study the magnetic correlation in the correlated paramagnetic regime in Ce$_2$Sn$_2$O$_7$. The observed diffuse scattering is inconsistent with AIAO order below $\mathrm{T_N}$ by exhibiting no short-range correlations at the AIAO wave-vectors, thus challenging the previously proposed nearest neighbor XYZ model. Instead, the diffuse scattering in Ce$_2$Sn$_2$O$_7$ strongly resembles the classical dipolar spin ice Dy$_2$Ti$_2$O$_7$, and suggests chains or clusters as the low-energy magnetic degrees of freedom. A chain-based order below $\mathrm{T_N}$ was argued to qualitatively account for the response of Ce$_2$Sn$_2$O$_7$ to a [1,1,0]-field.

\section*{Acknowledgements}
We acknowledge illuminating discussions with F. Desrochers, M.J.P. Gingras, Y.B. Kim, J.G. Rau, O. Benton, B. Placke, and Z. Shou. This work was supported in part by NSERC of Canada. The synthesis and crystal growth of the samples was performed at Clemson University supported by DOE grant DE-SC0020071. R. S. acknowledges support from the DFG under Project No. 575641691. A portion of this research used resources at the Spallation Neutron Source, a DOE Office of Science User Facility operated by the Oak Ridge National Laboratory, on proposal number IPTS-34379. Use of the MAD beamline at the McMaster Nuclear Reactor for crystal alignment is supported by McMaster University and the Canada Foundation for Innovation.
\section{Appendix}
\subsection{Comparison with Diffuse Scattering in Powder Ce$_2$Sn$_2$O$_7$}
\begin{figure}[tb]
\includegraphics[width=0.5\textwidth]{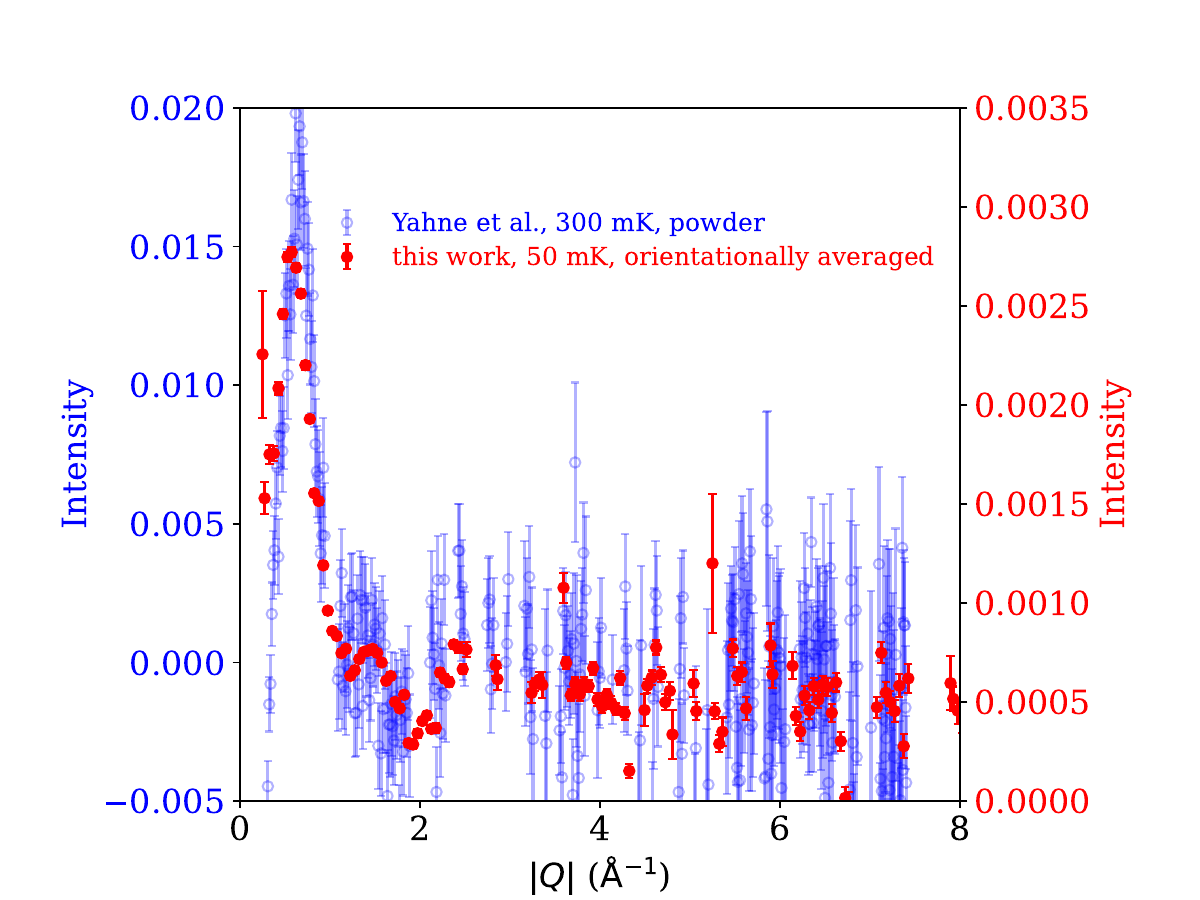}
\caption{\label{powdercompare}Comparison between the diffuse scattering in hydrothermally grown powder samples of Ce$_2$Sn$_2$O$_7$ \citep{Yahne_PRX_2024_Ce2Sn2O7} and the orientationally averaged single-crystal data (Fig.~\ref{largeQ}).}
\end{figure}
In Fig.~\ref{powdercompare}, we compare previous diffuse scattering measurements on hydrothermally grown powder samples of Ce$_2$Sn$_2$O$_7$ \citep{Yahne_PRX_2024_Ce2Sn2O7} with our orientationally averaged single-crystal data in Fig.~\ref{largeQ}. The quantitative agreement between the two measurements, performed on different sample batches and using different instruments, confirms the reproducibility of the observed low-$\mathbf{Q}$ diffuse scattering and the absence of a high-$\mathbf{Q}$ octupolar signal.

\subsection{Structural Defects in Ce$_2$Zr$_2$O$_7$}\label{disorder}
\begin{figure}[tb]
\includegraphics[width=0.5\textwidth]{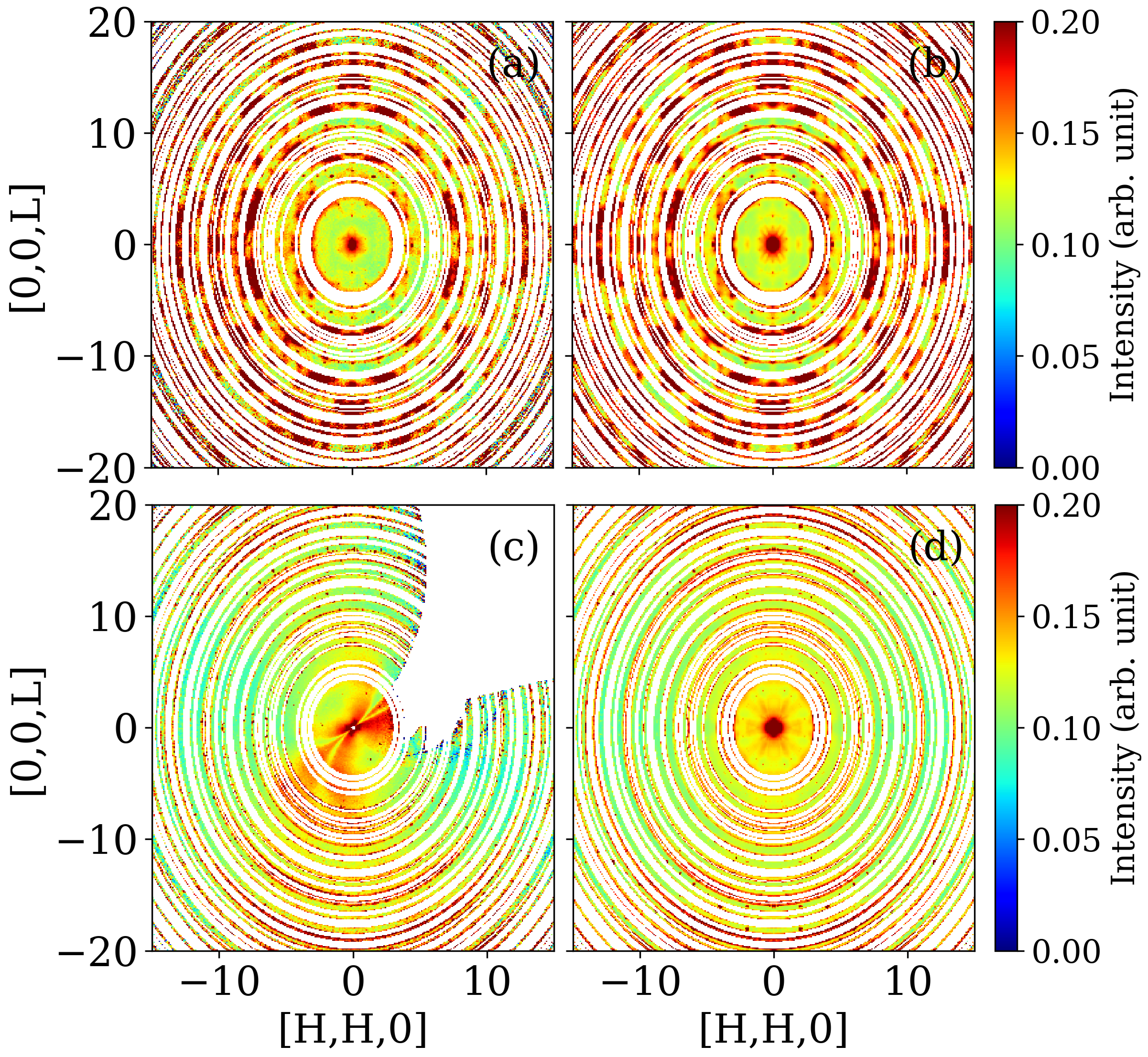}
\caption{\label{Ce2Zr2O7_defect} (Background-\textit{un}subtracted) diffuse scattering in the [H+0.5,H-0.5,L]-plane in (a,b) Ce$_2$Zr$_2$O$_7$ at 5~K and (c,d) Ce$_2$Sn$_2$O$_7$ at 12~K, obtained by integrating the diffuse intensity within [0.4,0.6]~r.l.u. along [H,-H,0]. (a,c) are unsymmetrized, while (b,d) are symmetrized according to the $Fd\bar{3}m$ space group. Powder rings from the copper mount and the sample environment have been removed from the data.}
\end{figure}

\begin{figure}[tb]
\includegraphics[width=0.5\textwidth]{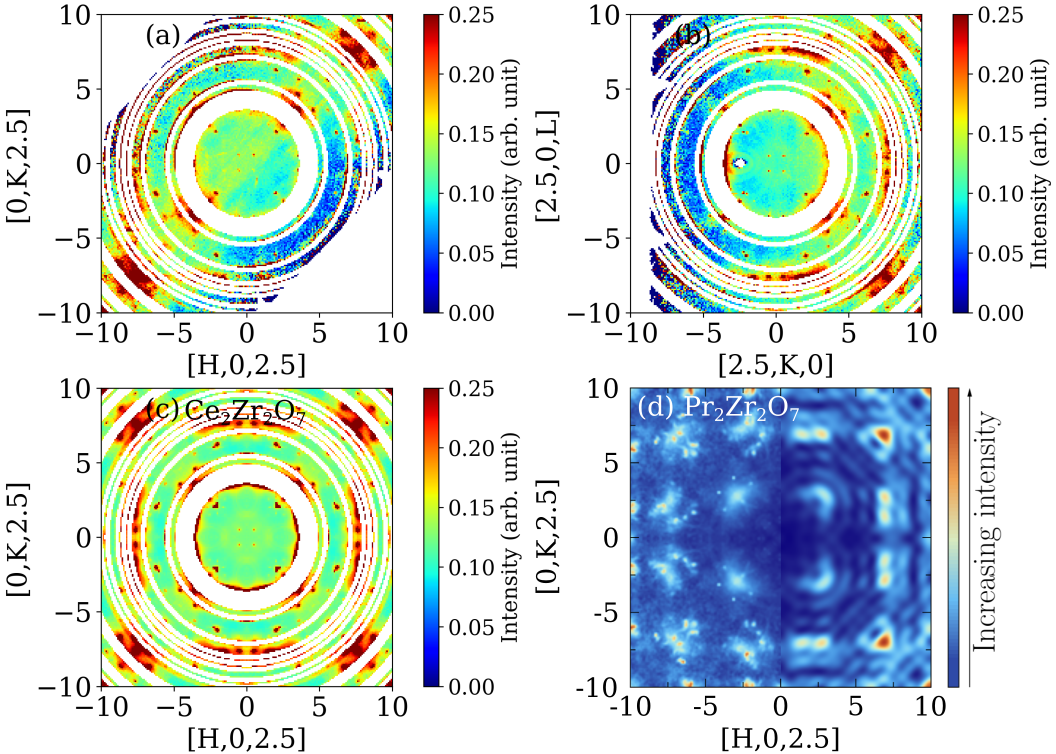}
\caption{\label{Ce2Zr2O7_Pr2Zr2O7} (a-b) Unsymmetrized diffuse scattering in the (a) [H,K,2.5] and (b) [2.5,K,L] plane at 5~K in Ce$_2$Zr$_2$O$_7$. The data after symmetrization with respect to the cubic space group is shown in (c). An integration range of $\pm 0.1~$r.l.u. perpendicular to the $\mathbf{Q}$ transfers of the data is used in (a-c). (d) Measured (left) and simulated (right) diffuse scattering in the [H,K,2.5] plane in Pr$_2$Zr$_2$O$_7$. Panel (d) of this figure is adapted from Ref.~\citep{Pr2Zr2O7_disorder_2025}.}
\end{figure}

The worse data quality in Ce$_2$Zr$_2$O$_7$ (See Supplemental Materials) might be related to a higher level of structural defects in the sample, as evidenced by the presence of structural diffuse scattering throughout the reciprocal space as shown in Fig.~\ref{Ce2Zr2O7_defect}~(a,b) and Fig.~\ref{Ce2Zr2O7_Pr2Zr2O7}~(a-c), for the [H+0.5, H-0.5, L] and [H,K,2.5]-planes, respectively. Unlike the weak temperature-dependent magnetic diffuse scattering that requires the subtraction of a high-temperature background, the observed structural diffuse scattering persists at high temperature (5~K in Ce$_2$Zr$_2$O$_7$) and is visible in the background-unsubtracted data. Its $\mathbf{Q}$-dependence clearly follows the lattice symmetry in the unsymmetrized data - as evidenced by the two-fold symmetric pattern in the [H+0.5, H-0.5, L]-plane [Fig.~\ref{Ce2Zr2O7_defect}~(a)] and almost identical diffuse scattering patterns in the symmetry-equivalent [H,K,2.5] [Fig.~\ref{Ce2Zr2O7_Pr2Zr2O7}~(a)] and [2.5,K,L]-planes [Fig.~\ref{Ce2Zr2O7_Pr2Zr2O7}~(b)], thus demonstrating its intrinsic origin. Symmetrized data in the [H+0.5, H-0.5, L] and [H,K,2.5] planes are shown in Fig.~\ref{Ce2Zr2O7_defect}~(b) and Fig.~\ref{Ce2Zr2O7_Pr2Zr2O7}~(c), respectively, which improves the signal-to-noise ratio of the diffuse scattering data without altering its $\mathbf{Q}$ dependence. Interestingly, we note a qualitative resemblance of the structural diffuse scattering in the [H,K,2.5] plane in Ce$_2$Zr$_2$O$_7$ [Fig.~\ref{Ce2Zr2O7_Pr2Zr2O7}~(c)] to that observed in Pr$_2$Zr$_2$O$_7$ [Fig.~\ref{Ce2Zr2O7_Pr2Zr2O7}~(d)] - both featuring strong intensities near (2.5, 2.5, 2.5) and (7.5, 7.5, 2.5), as well as a rod-like scattering along K connecting $(7.5,\pm 7.5, 2.5)$ [the rod-like feature is more pronounced in the simulated diffuse scattering in the right panel of Fig.~\ref{Ce2Zr2O7_Pr2Zr2O7}~(d)]. The $\mathbf{Q}$-dependence in Pr$_2$Zr$_2$O$_7$ could be reproduced by a defect structure where a small percentage of Pr occupies the Zr-site~\citep{Pr2Zr2O7_disorder_2025}. On the other hand, a quantitative modeling of the structural disorder in Ce$_2$Zr$_2$O$_7$ requires a better diffuse scattering measurement with less scattering by the sample environment and the sample holder, manifested in the present data (Fig.~\ref{Ce2Zr2O7_defect} and Fig.~\ref{Ce2Zr2O7_Pr2Zr2O7}) as densely populated powder rings.

In contrast, the diffuse scattering associated with structural defects is entirely absent in Ce$_2$Sn$_2$O$_7$ [Fig.~\ref{Ce2Zr2O7_defect}~(c,d)], directly confirming its high crystal quality. The `bow-tie'-shaped scattering near $\mathbf{Q}\approx 0$ in Fig.~\ref{Ce2Zr2O7_defect}~(c) can be attributed to attentuation by the copper plates (stronger when the copper plates are parallel to the incident and scattered beams) which, unlike the defect scattering in Ce$_2$Zr$_2$O$_7$, does not follow the crystal symmetry. The higher level of structural disorder in Ce$_2$Zr$_2$O$_7$ could arise from a higher growth temperature used in the floating zone method and/or a larger $\mathrm{Zr^{4+}}$, which could increase the A/B site mixing in the pyrochlore lattice.

\end{document}